\documentclass[9pt,shortpaper,twoside,web]{ieeecolor}
\pdfoutput=1
\usepackage{generic}
\usepackage{cite}
\usepackage{amsmath,amssymb,amsfonts}
\usepackage{algorithmic}
\usepackage{graphicx}
\usepackage{textcomp,nicefrac}
\usepackage[switch]{lineno}
\usepackage{booktabs}
\usepackage{float}
\usepackage{multirow}
\usepackage{etoolbox}
\usepackage{caption}
\usepackage{amsmath}

\UseRawInputEncoding
\makeatletter
\patchcmd{\@makecaption}
  {\scshape}
  {}
  {}
  {}
\makeatletter
\patchcmd{\@makecaption}
  {\\}
  {.\ }
  {}
  {}
\makeatother

\def\BibTeX{{\rm B\kern-.05em{\sc i\kern-.025em b}\kern-.08em
    T\kern-.1667em\lower.7ex\hbox{E}\kern-.125emX}}    
\begin{document}
\title{Stability of Charge Collection Efficiency and Time Resolution in a Novel Ultra-fast Graphene-Optimized Silicon Carbide Detector Under X-ray Irradiation}
\author{Zhenyu. Jiang, Congcong. Wang, Jingxuan. He, Yi Zhan, Yingjie Huang, Xiyuan. Zhang and Xin. Shi
\thanks{This work is supported by the National Natural Science Foundation of China (Nos. 12305207, 12375184 and 12405219) and support from CERN DRD3 Collaboration. (Corresponding author: Congcong Wang)}
\thanks{Zhenyu Jiang is with the Institute of High Energy Physics, Chinese Academy of Sciences, Beijing
100049, China,  and also with Liaoning University, Liaoning 110136, China.}
\thanks{Congcong Wang is with
the Institute of High Energy Physics, Chinese Academy of Sciences, Beijing
100049, China. (e-mail: wangcc@ihep.ac.cn).}
\thanks{Jingxuan He and Yi Zhan are with Qufu Normal University, Rizhao, 276827, China}
\thanks{Yingjie Huang is with Jilin University, Changchun, 130012, China}
\thanks{Xiyuan Zhang and Xin Shi are with the Institute of High Energy Physics, Chinese Academy of Sciences, Beijing 100049, China.}}
\maketitle

\begin{abstract}
A graphene-optimized silicon carbide PIN detector was fabricated and its radiation tolerance under X-ray irradiation of 160 keV was evaluated. Its electrical properties, charge collection performance and time resolution of $\beta$-particles ($^{90}$Sr) are reported. After 1 MGy irradiation, the detector maintains an ultralow leakage current of approximately 2.2~$\times$ 10$^{-10}$A @ 300 V and the C-V characteristics are basically consistent with full depletion at 120V. The time resolution of the graphene-optimized silicon carbide detector is 58.0~ps. The time resolution is comparable to that of state-of-the-art 4H-SiC low-gain avalanche detectors (LGADs). The G/RE 4H-SiC PIN detector exhibits outstanding time resolution performance. Compared with the time resolution of the RE 4H-SiC PIN detector, the time resolution of the G/RE 4H-SiC PIN detector has decreased by 39.6\%. This demonstrates the significance of the graphene electrode design. 
The graphene detector exhibits a charge collection efficiency (CCE) of 99.24\% after X-ray irradiation, along with excellent stability. The graphene-optimized silicon carbide detector maintains good timing resolution: 58.0 ps before and 64.0 ps after X-ray irradiation. Experimental results indicate that the CCE and time resolution performance exhibit good stability before and after irradiation.
These results demonstrate stable performance under extreme X-ray exposure, highlighting the detector’s potential for radiation-hard applications in high-energy physics, space missions, and nuclear reactor monitoring.

\end{abstract}


\section{Introduction}
\IEEEPARstart Silicon carbide (SiC) has advantages of high-temperature resistance, radiation hardness, low dark current, fast response time, high signal-to-noise ratio and high breakdown electric field. Consequently, SiC detectors have critical applications in aerospace, deep-space exploration, nuclear energy and high-energy physics, industrial and security monitoring, as well as ultraviolet detection. The presence of metal electrodes on the detector surface can absorb or scatter incident particles, leading to partial energy loss and affecting detection efficiency and energy resolution. However, the presence of metal electrodes on the detector surface can absorb or scatter incident particles, leading to partial energy loss and affecting detection efficiency and energy resolution. Neutron detection primarily relies on indirect methods through nuclear reactions with atomic nuclei to generate secondary charged particles. Metal electrodes may absorb these secondary charged particles, such as alpha particles and recoil protons, resulting in signal loss or attenuation\cite{1611007}. Low-energy soft X-rays and gamma rays can be partially absorbed within the metal electrode layer through processes like the photoelectric effect and Compton scattering, thereby reducing detection efficiency\cite{10.1117/12.679554}. High-energy particles striking the metal electrodes may produce secondary particles, such as delta electrons and bremsstrahlung X-rays\cite{PhysRev.96.1199}. These secondary particles can enter the sensitive region of the detector, generating false signals, increasing background noise, and interfering with the identification of true target events. Graphene possesses unparalleled thinness, excellent electrical conductivity and extremely low mass. Graphene, as a transparent electrode, not only provides excellent channels for charge collection and transmission, but also minimizes the interference of the incident signal to the lowest theoretical limit. The most important thing is that graphene can enhance the charge collection stability and temporal resolution of radiation detectors. This avoids the instability in charge collection and the decline in timing performance caused by traditional electrode windowing. Fast-timing detectors are utilized to suppress pile-up events and improve signal-to-noise ratios in high-luminosity collider environments in high-energy physics; to aid tracking detectors in resolving decay trajectories in space exploration; to enable particle identification and neutron imaging in nuclear physics; and to facilitate precise proton therapy in medical physics. 

In high-intensity irradiation scenarios such as medical imaging, security screening, deep-space exploration and scientific research, people are increasingly emphasizing the radiation hardness of X-ray detectors to ensure long-term stable operation, reduce the risk of failure, and meet the demands of lower-dose imaging and more extreme environmental applications\cite{radiation1030018,  10.1063/1.4894019,Ryan_2018}. In nuclear reactors, fuel cycle facilities, and waste storage sites, continuous monitoring of X-rays is essential for personnel safety and structural integrity assessment\cite{viswanathan2018quantitative,Ramli_2020}. In high-energy physics experiments, detectors on particle colliders and accelerators are exposed to extreme radiation doses while needing to accurately record and measure the energy of X-rays from collision events\cite{s23177328,20220205,Lee2014RadiationTF,10.1063/5.0040571,FROJDH201543 }. In the medical fields of radiotherapy and advanced imaging, precise real-time dosimetry is crucial for maximizing tumor control while minimizing damage to healthy tissues\cite{10.3389/fphy.2025.1576227,KEALL2025787,FREITASNASCIMENTO2025107344,https://doi.org/10.1118/1.4926282,Jegal2024MotionMA}. In space exploration, satellites and probes are continuously bombarded by cosmic rays and solar X-rays, necessitating detectors for navigation, scientific instrumentation, and monitoring electronic systems against single-event effects (SEE) and total ionizing dose (TID) transients\cite{20220205,PALMERINI2002159}. All these requirements demand that detectors possess radiation hardness, high signal-to-noise ratio, high sensitivity, and fast time response, without degradation due to cumulative effects, while operating stably and providing accurate measurement results. Conventional silicon detectors exposed to high-energy X-ray irradiation exhibit pronounced susceptibility to TID effects and displacement damage (DD), which induce a marked rise in leakage current, deterioration of the CCE, and ultimately device failure. Consequently, this severely constrains their deployment in extreme radiation environments. The significance of studying the CCE and time resolution properties of graphene-optimized silicon carbide PIN detector under high-intensity X-ray irradiation lies in validating its ability to maintain core functionalities under extreme radiation environments.

In this work, the ring electrode (RE) 4H-SiC PIN and graphene-optimized 4H-SiC PIN detectors before and after X-ray irradiation with an energy of 1 MGy the have been fabricated. Their electrical characteristics were systematically evaluated in current-voltage (I-V) and capacitance-voltage (C-V) curves. The differences in CCE and time resolution between the RE 4H-SiC PIN and the G/RE 4H-SiC PIN were compared and analyzed, demonstrating the effectiveness of graphene in optimizing time resolution. The CCE and time resolution properties of the G/RE 4H-SiC PIN detector under an X-ray irradiation energy of 1 MGy were analyzed, and the stability of the device was verified. 

\section{Detector fabrication and irradiation conditions}

\subsection{Epitaxial structure and detector fabrication}

The SiC PIN detector employs a fully epitaxial vertical PIN architecture, the epitaxial structure  from bottom to top includes:

1) The conductive N-type 4H-SiC substrate with a thickness of 350~$\mu$m.

2) The lightly doped N-epi layer with a nitrogen ion doping concentration of 5~ $\times$ 10$^{13}$ cm$^{-3}$ and a thickness of 50~$\mu$m.

3) The P++ layer with an aluminum ion doping concentration of 2~$\times$ 10$^{19}$ cm$^{-3}$ and a thickness of 0.6 $\mu$m. 

\begin{figure}[htbp]
\centering
\includegraphics[scale=1]{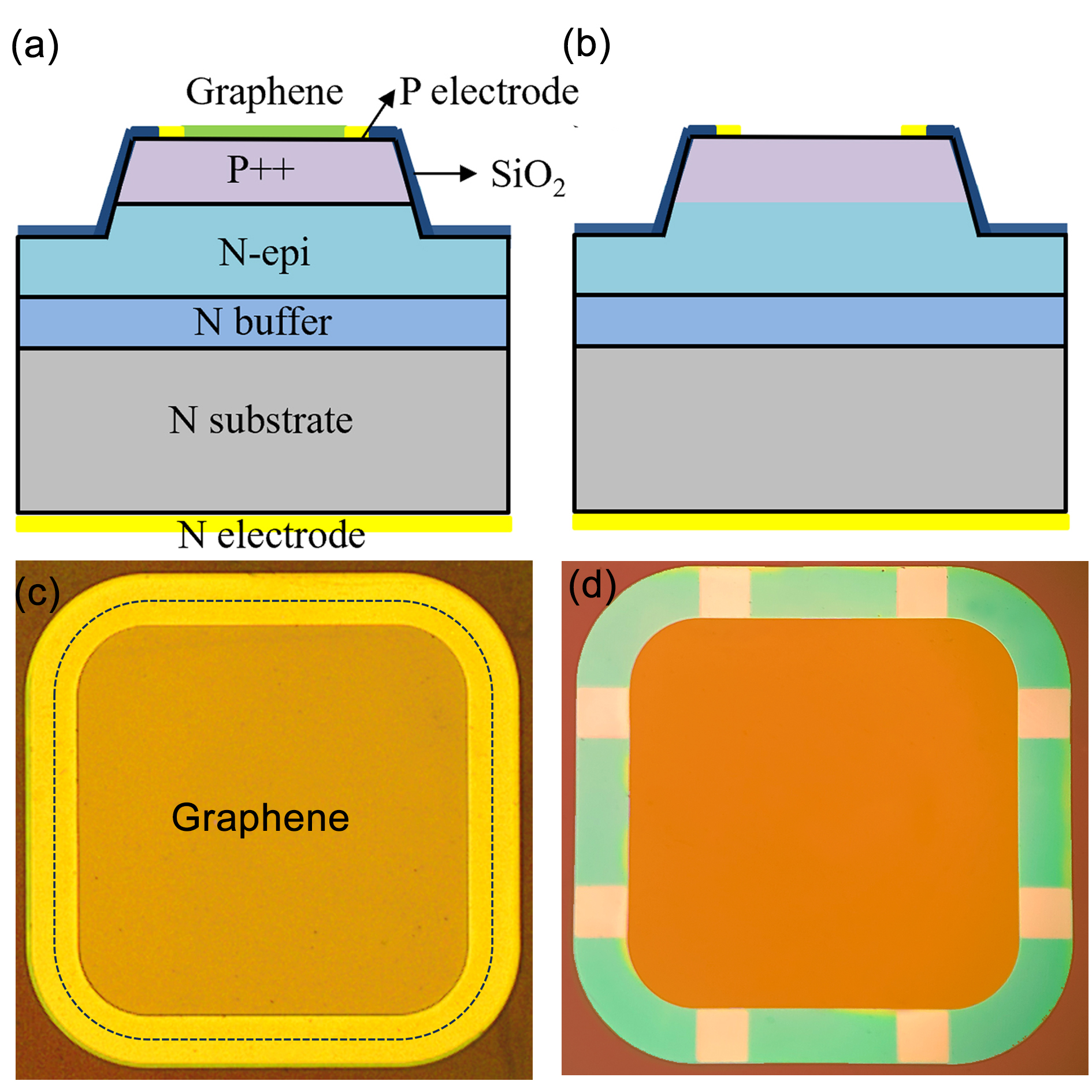}
\caption{(a) Cross-sectional structure of the G/RE 4H-SiC PIN detector. (b) Cross-sectional structure of the RE 4H-SiC PIN detector. (c) Real G/RE 4H-SiC PIN detector image. ( The area within the black dashed line contains graphene. ) (d) Real RE 4H-SiC PIN detector image.}
\label{figure}
\end{figure}

The RE and G/RE 4H-SiC PIN detectors were fabricated. Fig.~1~(a-d) show the cross-sectional structures and  real detector image of the RE 4H-SiC PIN detector and G/RE 4H-SiC PIN detector. The structure of the G/RE 4H-SiC PIN includes monolayer graphene, P electrode, SiO$_{2}$ passivation layer, P++ layer, N-epi layer, N buffer layer, conductive N-type 4H-SiC substrate and N electrode as shown in Fig.~1 (a). The fabrication process of the G/RE 4H-SiC PIN detector mainly includes lithography, etching, electron beam evaporation, magnetron sputtering, rapid thermal annealing, transfer and etch graphene. Etching depth of epitaxial structure is more than 0.6~$\mu$m to ensure full etching of P++ layer. The Ni/Ti/Al = 50nm/15nm/60nm as the electrode was grown on the top of P++ layer and N-type substrate by using electron evaporating method. The rapid annealing time and temperature are 2 min and 950~$^\circ$C to form P-ohmic contact. A SiO$_{2}$ layer with a thickness of 500 nm was deposited by plasma-enhanced chemical vapor deposition (PECVD) at 350~$^\circ$C. The size of the detectors is 2mm~$\times$ 2mm. Monolayer graphene (SixCarbon Technology) on a copper substrate is transferred to the surface of a silicon carbide detector by wet transfer method. After the graphene transfer is completed, the graphene graphics are realized by photolithography and reactive ion etching (RIE).

\subsection{Irradiation conditions}

X-ray irradiation was conducted using an X-ray irradiation facility at the Institute of High Energy Physics (IHEP), Chinese Academy of Sciences. The irradiation energy was 160 keV X-ray beam at a dose rate of 246 Gy/min (silicon equivalent), and the irradiation dose was 1 MGy. The irradiation temperature was room temperature, and the devices were unbiased during irradiation.

\section{ELECTRICAL PERFORMANCE ANALYSIS}

The I-V characteristics of the RE, unirradiated G/RE and X-ray irradiated G/RE 4H-SiC PIN detectors are shown in Fig. 2~(a). At a reverse bias voltage of 300 V, the leakage currents of these detectors remain effectively suppressed at an extremely low level of  1~$\times$ 10$^{-10}$~A. The leakage current curves before and after irradiation almost coincide, indicating no increase in current due to X-ray radiation damage. This excellent electrical stability stands in sharp contrast to conventional silicon-based detectors. Under the same irradiation conditions, the leakage current in silicon detectors typically increases by several orders of magnitude due to a substantial rise in both bulk and surface leakage currents.\cite{Zhang_2011}. There are two reasons for the relatively low leakage current of the silicon carbide detector before and after irradiation. The reverse leakage current of silicon carbide is extremely low, which is mainly attributed to its wide bandgap property. The wide bandgap significantly reduces the intrinsic carrier concentration, thereby exponentially suppressing the leakage current dominated by thermal excitation\cite{SINGH2006713}. More importantly, X-rays primarily deposit ionization energy in matter, resulting in extremely low non-ionizing energy loss (displacement damage). Silicon carbide, owing to its high atomic bond energy, possesses a higher displacement damage threshold, thereby further suppressing the formation of defects responsible for leakage current generation\cite{COWEN201873}.

The C-V characteristics of the RE, unirradiated G/RE and irradiated G/RE 4H-SiC PIN detectors are shown in Fig. 2~(b).The C-V curves show that the full depletion voltage of these detectors is about 120~V. The effective doping concentration and depletion depth of N-epi layer are about 4.5~$\times$ 10$^{13}$~cm$^{-3}$ and 45~$\mu$m, respectively. The experiment proved that the total depletion voltage and the effective doping concentration remained basically unchanged before and after irradiation. The fundamental reason is that X-rays primarily cause ionization energy loss, and their non-ionizing energy loss (displacement damage) is extremely low, insufficient to introduce significant lattice defects into the semiconductor bulk material\cite{4033220}. Since the bulk defect density does not increase significantly, the effective doping concentration remains unchanged, and consequently the full depletion voltage shows no significant variation\cite{10.1063/5.0179556}. The effective doping concentration approaching the lowest doping level in SiC epitaxial growth technique. The doping concentration and thickness meet the design requirements. It can be seen from the Fig.~2~(b), whether the graphene is irradiated or not basically does not affect the electrical performance of the detectors.

\begin{figure}[htbp]
\centering
\includegraphics[scale=0.9]{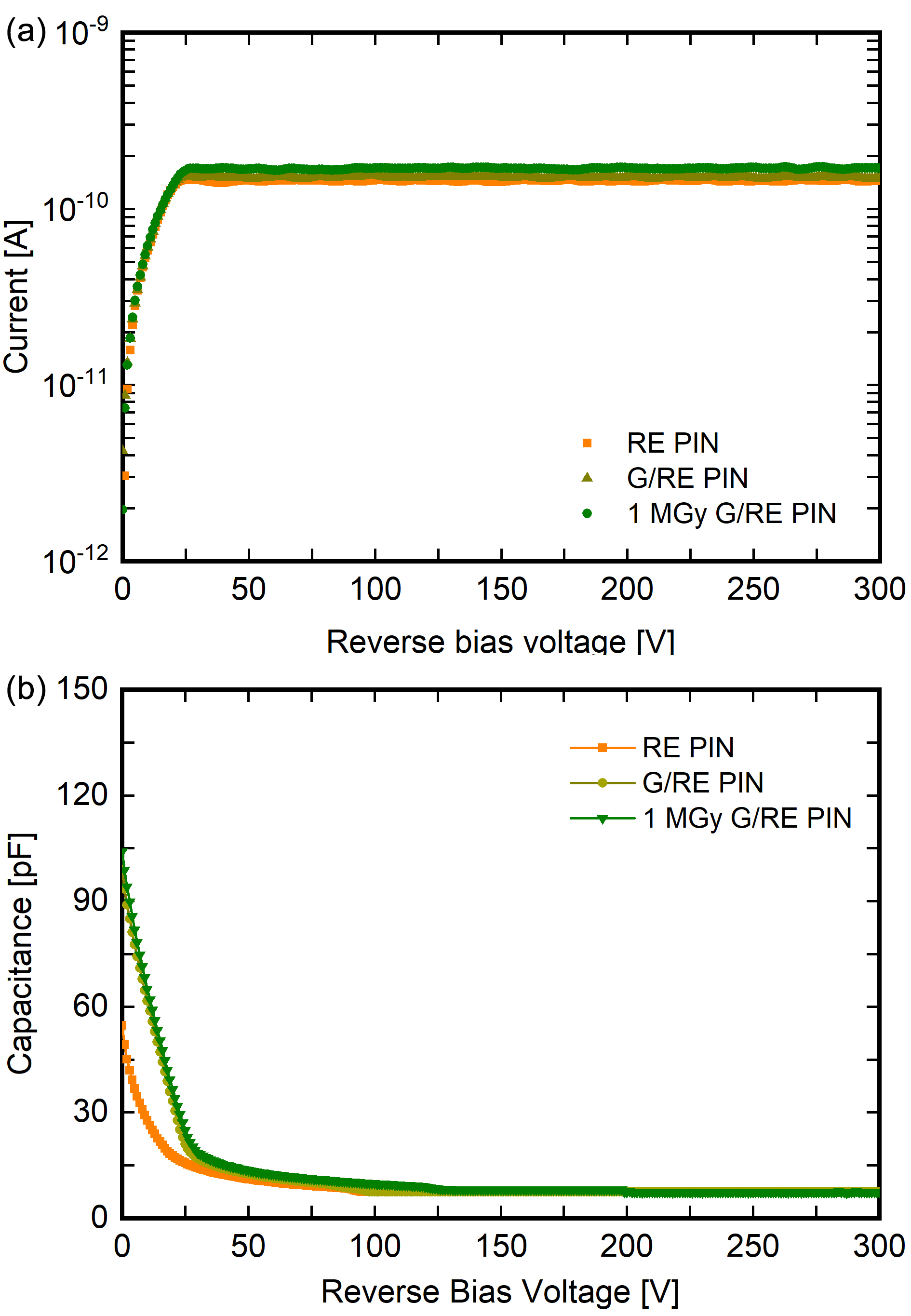}
\caption{The RE 4H-SiC PIN, unirradiated G/RE 4H-SiC PIN detectors and 1MGy irradiated G/RE 4H-SiC PIN detectors: (a) I-V characteristics. (b) C-V characteristics.}
\label{figure}
\end{figure}

\section{Charge Collection Efficiency Performance}

The charge collection performance setup for $\beta$ particles is shown in Fig. 3. The system includes $^{90}$Sr radioactive source, 4H-SiC PIN, single channel electronic readout board, high voltage source (Keithley 2470), low voltage source (GPD-3303, SGWINSTE) and oscilloscope (MSO64, Tektronix 2.5 GHz). The detectors are encapsulated on the electronic readout board by using conductive adhesive, and the pad electrode of detectors are connected to the readout board. The high voltage source provides the detector with reverse bias. The low voltage source provides power supply to the single channel electronic readout board.

\begin{figure}[htbp]
\centering
\includegraphics[scale=1]{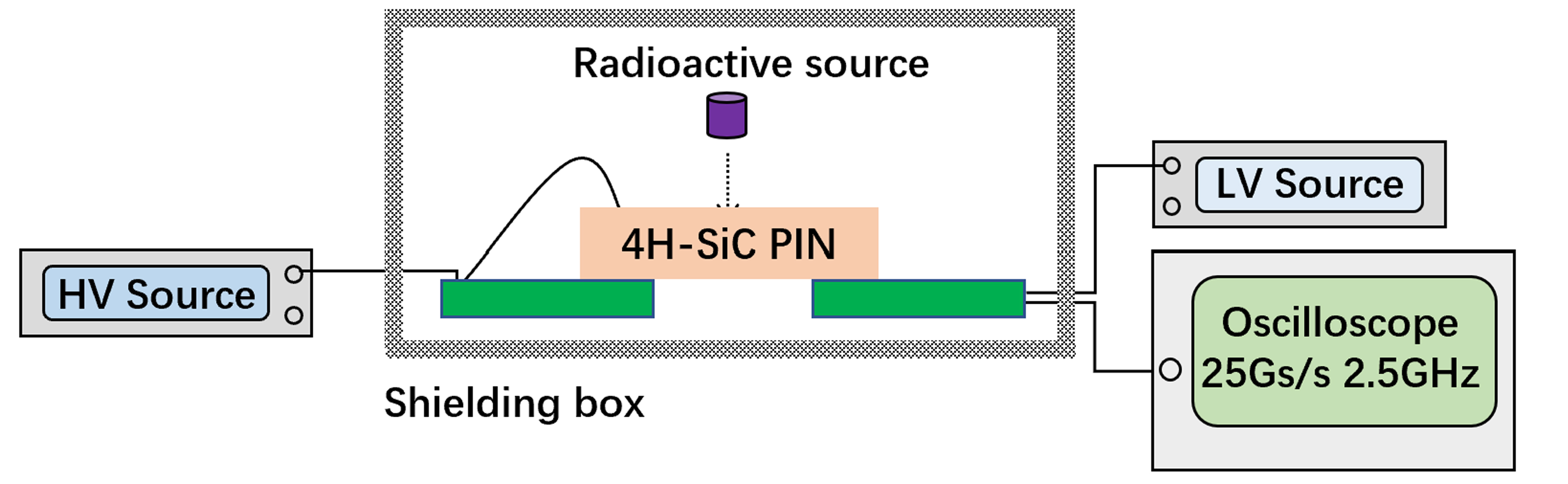}
\caption{Experimental setup for charge collection efficiency of a $^{90}$Sr source.}
\label{figure}
\end{figure}

The $\beta$ particle signals, which follow a Landau distribution due to their energy loss straggling, were fitted with a Landau function as shown in Fig.~4~(a). The most probable value (MPV) of each fit was taken as the characteristic collected charge under the corresponding condition. The Landau distribution, with its asymmetric high-energy tail, offers a more physically realistic description of energy deposition in thin detectors than the symmetric Gaussian distribution. Fig.~4~(b) presents the CCE as a function of reverse bias voltage for these detectors. The full depletion voltages are about 120 V. For voltages greater than 120 V no further increase of the depletion depth occurs, and no increase in the collected charge is observed.The collected charges of the RE 4H-SiC PIN, the unirradiated G/RE 4H-SiC PIN and X-ray irradiated G/RE 4H-SiC PIN detectors are 1.74 fC, 1.76 fC and 1.64~fC at 300 V. The charge collection efficiency (CCE) of the unirradiated G/RE 4H-SiC PIN detector was defined as 100\% @ 300 V. Under identical bias conditions, the RE detector and the X-ray irradiated G/RE 4H-SiC PIN detector exhibited CCEs of 97.99\% and 99.24\%, respectively. It demonstrates that the 4H-SiC PIN detector maintains excellent charge-collection performance for $\beta$ particles even under extremely high X-ray doses. This result fully highlights the significant performance advantages of the wide-bandgap semiconductor silicon carbide in harsh radiation environments.

\begin{figure}[htbp]
	\centering
	\includegraphics[scale=1]{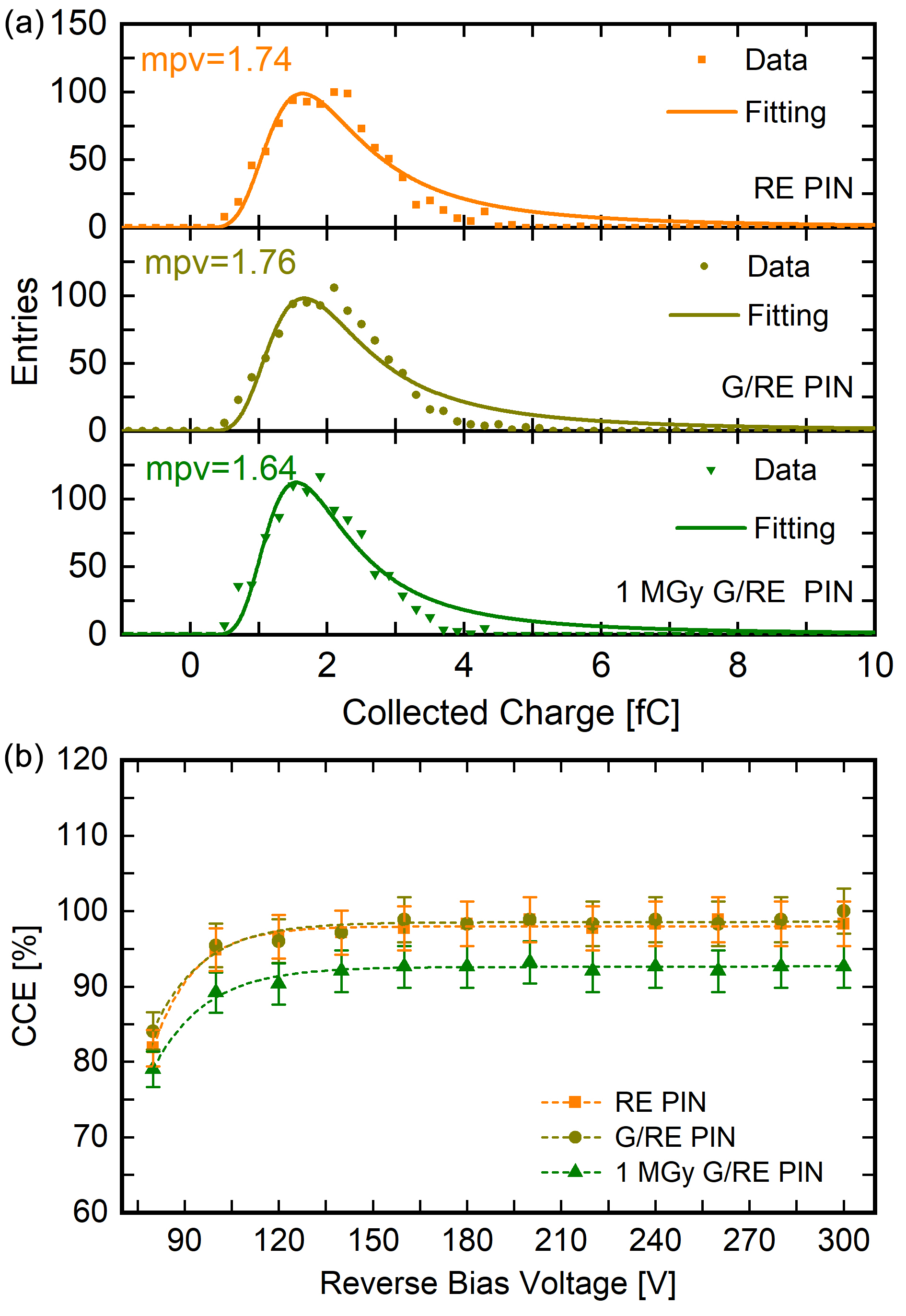}
	\caption{Charge collection performance of 4H-SiC PIN detectors at 300V: (a) Landau fit of the collected charge spectrum, with MPV indicating characteristic charge of detectors. (b) CCE versus reverse bias for the RE, unirradiated G/RE and 1MGy X-ray irradiated G/RE 4H-SiC PIN}
	\label{figure}
\end{figure}

\section{Time Resolution Performance}

The time resolution setup for $\beta$ particles is shown in Fig.~5. This measurement system adopts a dual-channel configuration for time resolution measurements. The reference channel (REF) employs a silicon low-gain avalanche detector (Si-LGAD, IHEP) with a 37~ps time resolution @ 200V. The other channel is the device under test (DUT) channel, which is used to mount the 4H-SiC device to be evaluated for its time resolution. Both detectors were exposed to $\beta$-particles from the $^{90}$Sr source. The signals from both channels were read out using identical UCSC boards followed by PE15A1008 main amplifiers (20 dB gain).A hole with a diameter of approximately 3 mm was created at the center of the bottom of the Si-LGAD readout board to ensure the passage of β particles. A high-voltage source is provided for each of the two channels to apply reverse bias. When an incident particle passes through the detector, the resulting current signal is first amplified by the pre-amplifier on the readout board.The signal is then sent to a 20 dB main amplifier for further amplification. The signals from both channels are acquired by a high-speed oscilloscope with a single-channel sampling rate of 25 Gs/s. The generated synchronous pulse trigger signal was acquired using an oscilloscope, and the jitter of the trigger was negligible.

\begin{figure}[htbp]
\centering
\includegraphics[scale=0.63]{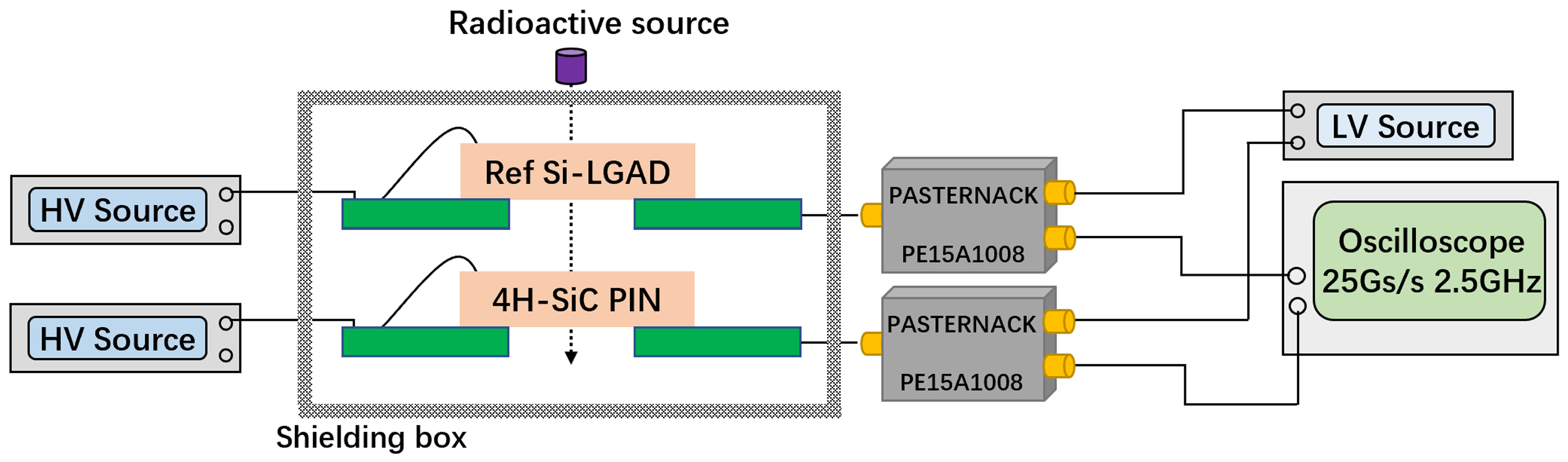}
\caption{Experimental setup for time resolutin of a $^{90}$Sr source.}
\label{figure}
\end{figure}

Fig. 6 shows the signal waveforms of Si LGAD and 4H-SiC PIN. The time difference between the DUT signal and REF signal is defined as $\Delta$t = $\text{t}_{\text {DUT}}$-$\text{t}_{\text {Ref}}$, where $\text{t}_{\text {DUT}}$ and $\text{t}_{\text {Ref}}$ are the arrival times of the Si-LGAD signal and the PIN signal, respectively. The time resolution is closely related to the probability distribution of the time difference Δt. To mitigate the influence of time walk—the variation in measured arrival time caused by differing signal amplitudes—on the time resolution, this paper employs the Constant Fraction Discrimination (CFD) method for data processing. The CFD method determines the trigger time at a constant fraction (50\%) of the signal amplitude, making the timing independent of pulse height. The process involves:

1) Finding the maximum amplitude Vmax and its time tmax.

2) Setting the threshold at Vth = 0.5 × Vmax.

3) Locating the point where the rising edge first crosses Vth and interpolating to obtain the precise CFD time.

The time spread $\sigma_{\text {$\Delta$t}}$ was extracted by performing a Gaussian fit to
the distribution of $\Delta$t. The timing resolution of the device under test, σDUT, was then calculated using:

\begin{equation}
\sigma_{D U T}=\sqrt{\sigma_{\Delta t}^{2}-\sigma_{Ref}^{2}}
\end{equation}

where $\sigma_{\text {Ref}}$ is 37.0 ps is the known resolution of the reference Si-LGAD detector.

\begin{figure}[htbp]
\centering
\includegraphics[scale=0.8]{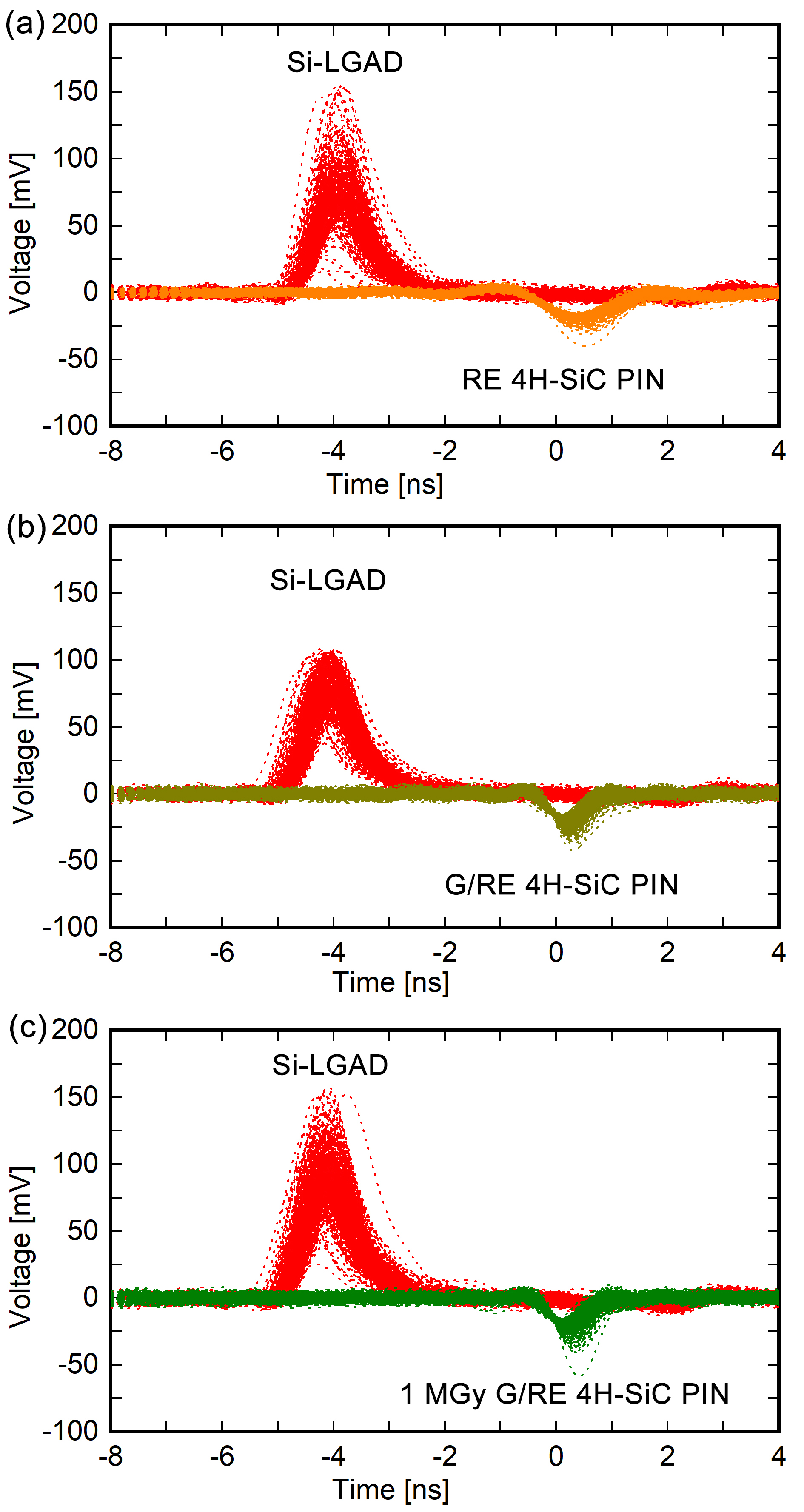}
\caption{Signal waveforms: (a) Si LGAD and RE 4H-SiC PIN detectors. (b) Si LGAD and G/RE 4H-SiC PIN detectors. (c) Si LGAD and 1MGy X-ray irradiated G/RE 4H-SiC PIN detectors.}
\label{figure}
\end{figure}

Fig. 7 shows the timing performance of the RE 4H-SiC PIN detector, unirradiated G/RE 4H-SiC PIN detector and irradiated G/RE 4H-SiC PIN detector at 300V bias. As shown in Fig. 7~(a), the G/RE 4H-SiC PIN detector exhibits excellent time resolution performance, with a value of 58.0 ps. The time resolution is comparable to that of state-of-the-art 4H-SiC low-gain avalanche detectors (LGADs)\cite{11303914}. The G/RE 4H-SiC PIN detector exhibits outstanding time resolution performance. As shown in Fig.~7~(b), the time resolution of the RE 4H-SiC PIN detector is 96.0 ps. Compared with the time resolution of the RE 4H-SiC PIN detector, the time resolution of the G/RE 4H-SiC PIN detector has decreased by 39.6\%. This demonstrates the significance of the graphene electrode design. As shown in Fig.~7~(c), after 1 MGy X-ray irradiation, the G/RE 4H-SiC PIN detector achieves a time resolution of 64.0~ps. This proves that under high-dose radiation conditions, the G/RE 4H-SiC PIN detector also exhibits an exceptionally high time resolution.

\begin{figure}[htbp]
\centering
\includegraphics[scale=0.9]{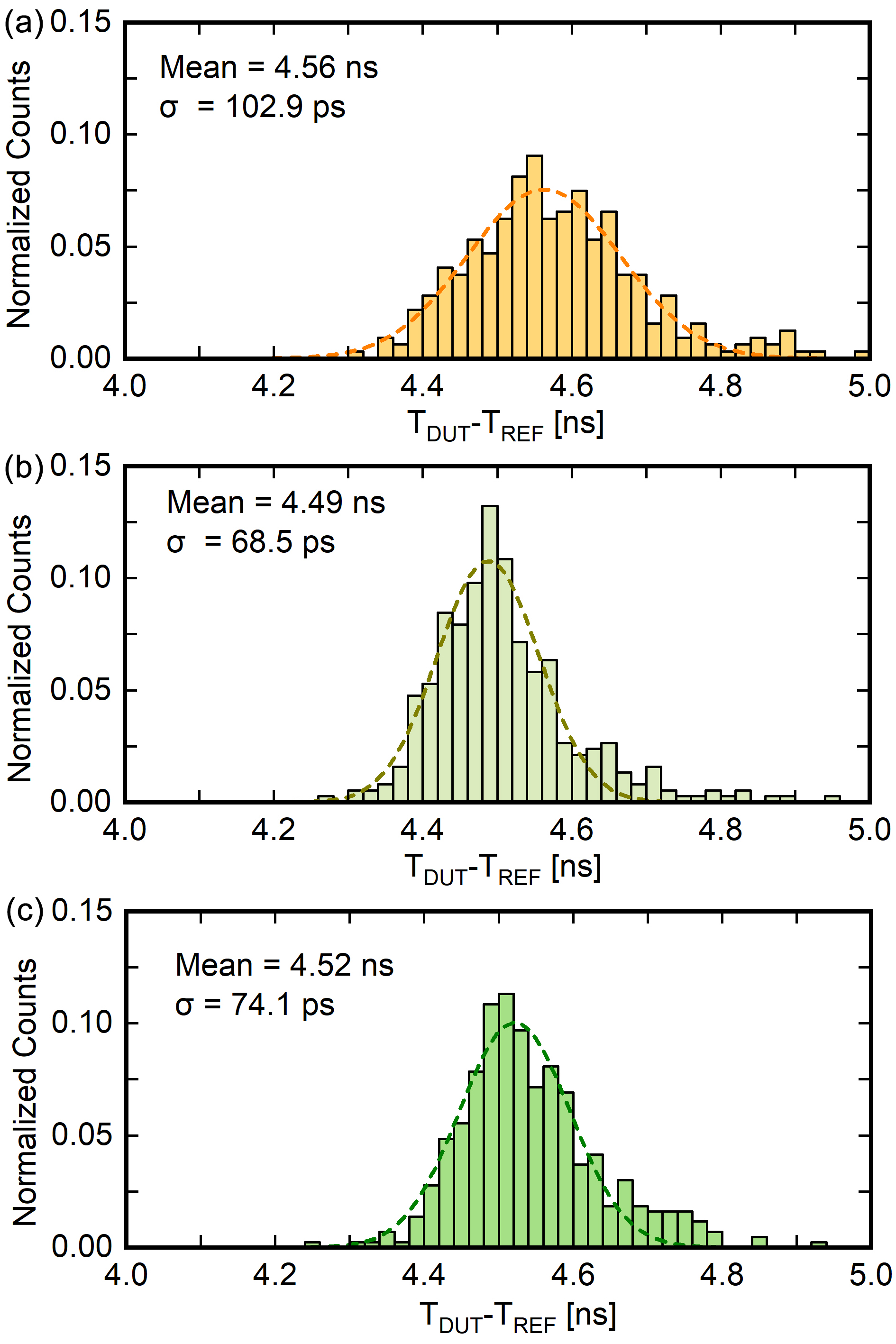}
\caption{Timing performance at 300V bias: (a) RE 4H-SiC PIN detector. (b) G/RE 4H-SiC PIN detector. (c) 1MGy X-ray irradiated G/RE 4H-SiC PIN detector.}
\label{figure}
\end{figure}

The time resolution $\sigma_{\text {Total}}$ of the detector consists of the following components:

\begin{equation}
\sigma_{Total}^{2}=\sigma_{TW }^{2}+\sigma_{T D C}^{2}+\sigma_{\text {Jitter }}^{2}
\end{equation}

where $\sigma_{\text{TW}}$ represents the time walk, which has been eliminated using CFD method. $\sigma_{\text{TDC}}$ is an error introduced by an uncertain shift in timing caused by analogue-to-digital conversion. However, as this experiment utilises an oscilloscope with a high sampling rate of 25~Gs/s, this error is negligible. $\sigma_{\text{Jitter}}$ is the main contributor to $\sigma_{\text {Total}}$ which arises from noise in the detector and front-end electronics and can be estimated as:

\begin{equation}
\sigma_{\text {Jitter }}=\frac{N}{d V / d t} \simeq=\frac{T_{\text {rise }}}{S / N}    
\end{equation}

where N is the noise, dV/dt is the rate of rise of the signal at the threshold, 
$\text{T}_{\text{rise}}$ is the rise time of the signal and S/N is the signal-to-noise ratio. This equation shows that the larger signal amplitude and faster rise time can result in smaller Jitter. Fig. 8 shows the Jitter distribution of the RE 4H-SiC PIN detector, unirradiated G/RE 4H-SiC PIN detector and irradiated G/RE 4H-SiC PIN detector. $\sigma_{\text{Jitter}}$ increased from 40.5 ps to 41.4 ps, primarily due to a reduction in the signal-to-noise ratio. The rise time distribution remained stable, indicating that the carrier transport characteristics did not deteriorate significantly as a result of irradiation.

\section{Discussion}

The experimental results have shown that no significant performance degradation was observed in the G/RE 4H-SiC PIN detector following exposure to 1 MGy of X-ray irradiation, confirming its excellent radiation resistance. The 160 keV X-ray irradiation produces secondary electrons with maximum energy of 120 keV. For a head-on collision with a Si atom, the maximum transferable energy is approximately 9eV and similarly $\sim$21 eV for C atoms ($M_C \approx 12m_p$). Both values are below the displacement threshold energy of SiC ($E_d \approx$ 25$\sim$35~eV for Si, 20$\sim$25~eV for C)\cite{10.1063/1.364397}. The maximum kinetic energy transferable to a lattice atom through elastic collision is severely limited by conservation of momentum. Furthermore, high-energy electrons predominantly lose energy through electronic stopping (ionization) rather than nuclear collisions, resulting in a Non-Ionizing Energy Loss (NIEL) approximately four orders of magnitude lower than that of MeV-scale protons. These secondary electrons cannot generate bulk displacement defects such as Z$_{1/2}$ or EH3 centers. The increased leakage current is therefore attributed to changes in the graphene electrode properties rather than bulk compensation effects, consistent with reported X-ray irradiation behavior in SiC detectors. The leakage current of the G/RE 4H-SiC PIN detector is less than 0.2 nA after 1~MGy irradiation, which demonstrates the excellent electrical stability of the detector. Finally, the high carrier saturation drift velocity of SiC ensures that the detector maintains stable and rapid charge collection and signal response capabilities even in environments with elevated defect concentrations. Consequently, the irradiated G/RE 4H-SiC PIN detector demonstrates excellent performance in terms of electrical characteristics, charge collection efficiency, and time resolution.

\begin{figure}[htbp]
\centering
\includegraphics[scale=0.9]{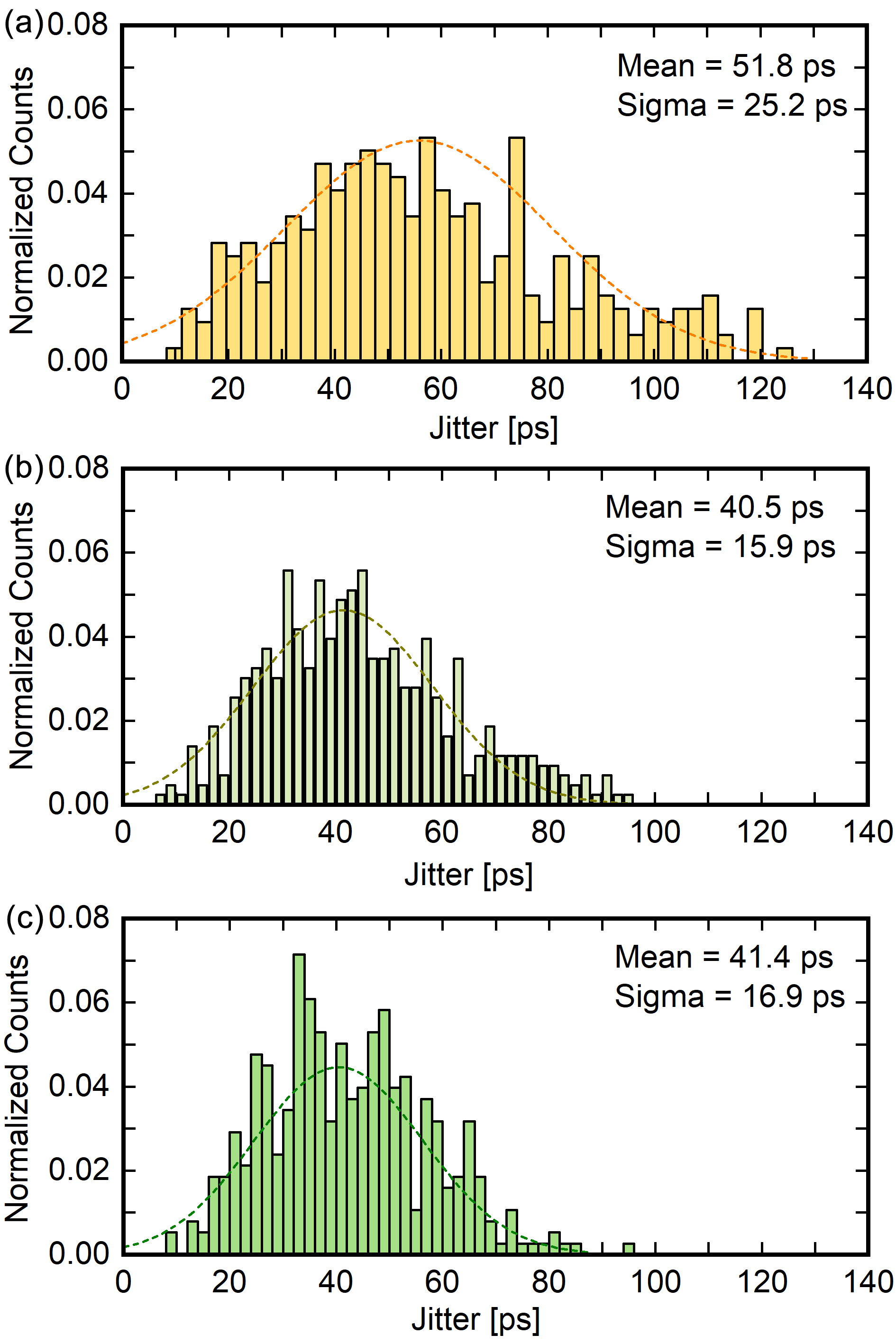}
\caption{Jitter distribution: (a) RE 4H-SiC PIN detector. (b) G/RE 4H-SiC PIN detector. (c) 1MGy X-ray irradiated G/RE 4H-SiC PIN detector.}
\label{figure}
\end{figure}

\section{Conclusion}
A graphene-optimized silicon carbide PIN detector was fabricated and its radiation tolerance under X-ray irradiation of 160 keV was evaluated. Its electrical properties, charge collection performance and time resolution of $\beta$-particles ($^{90}$Sr) are reported. After 1 MGy irradiation, the detector maintains an ultralow leakage current of approximately 2.2~$\times$ 10$^{-10}$ A @ 300 V and the C-V characteristics are basically consistent with full depletion at 120V. The time resolution of the graphene-optimized silicon carbide detector is 58.0~ps. The time resolution is comparable to that of state-of-the-art 4H-SiC low-gain avalanche detectors (LGADs). The G/RE 4H-SiC PIN detector exhibits outstanding time resolution performance. Compared with the time resolution of the RE 4H-SiC PIN detector, the time resolution of the G/RE 4H-SiC PIN detector has decreased by 39.6\%. The graphene detector exhibits a charge collection efficiency (CCE) of 99.24\% after X-ray irradiation, along with excellent stability. This demonstrates the significance of the graphene electrode design. The graphene-optimized silicon carbide detector maintains good timing resolution: 58.0 ps before and 64.0 ps after X-ray irradiation. Experimental results indicate that the CCE and time resolution performance exhibit good stability before and after irradiation. These results demonstrate stable performance under extreme X-ray exposure, highlighting the detector’s potential for radiation-hard applications in high-energy physics, space missions, and nuclear reactor monitoring.

Our team has successfully fabricated graphene-optimized SiC PiN and SiC LGAD detectors. In addition, we will continue to advance the research and development of next-generation devices, including SiC AC-LGAD, SiC BJT, and SiC DC-RSD. Among these, the SiC AC-LGAD enables simultaneous measurement of submillimeter spatial resolution and picosecond temporal resolution; the SiC BJT explores potential applications in detector readout electronics and signal amplification; and the SiC DC-RSD combines excellent radiation tolerance with fast temporal response. The development of these novel devices will further enrich the SiC detector technology portfolio and promote their engineering applications in four-dimensional tracking, time-resolved imaging, and extreme radiation environments.

\section{Acknowledgement}
We acknowledge the RASER team (https://raser.team) and CERN DRD3 Collaboration (https://drd3.web.cern.ch) for their useful discussions.

\bibliographystyle{ieeetran}
\bibliography{sample}

@ARTICLE{1611007,
  author={Zahraman, K. and Roumie, M. and Raulo, A. and Auricchio, N. and Ayoub, M. and Donati, A. and Dusi, W. and Hage-Ali, M. and Lmai, F. and Perillo, E. and Siffert, P. and Sowinska, M. and Ventura, G.},
  journal={IEEE Transactions on Nuclear Science}, 
  title={{Study of the thickness of the dead Layer below electrodes, deposited by electroless technique, in CdTe nuclear detectors}}, 
  year={2006},
  volume={53},
  number={1},
  pages={378-382},
  doi={10.1109/TNS.2006.869846}}

@inproceedings{10.1117/12.679554,
author = {P. Paki Amouzou and M. Gertsenshteyn and T. Jannson and P. Shnitser and G. Savant},
title = {{Inelastic scattering measurements of low-energy x-ray photons by organics, soil, water, and metals}},
volume = {6319},
booktitle = {Hard X-Ray and Gamma-Ray Detector Physics and Penetrating Radiation Systems VIII},
editor = {F. Patrick Doty and H. Bradford Barber and Hans Roehrig and Larry A. Franks and Arnold Burger and Ralph B. James},
publisher = {SPIE},
pages = {63190V},
year = {2006},
doi = {10.1117/12.679554},

}

@article{PhysRev.96.1199,
  title = {{Emission of Energetic Secondary Electrons Produced by 1.3-Mev Electron Bombardment}},
  author = {Shatas, R. A. and Marshali, J. F. and Pomerantz, M. A.},
  journal = {Phys. Rev.},
  volume = {96},
  issue = {5},
  pages = {1199--1203},
  numpages = {0},
  year = {1954},
  month = {Dec},
  publisher = {American Physical Society},
  doi = {10.1103/PhysRev.96.1199},
  
}

@Article{radiation1030018,
AUTHOR = {Karmakar, Arijit and Wang, Jialei and Prinzie, Jeffrey and De Smedt, Valentijn and Leroux, Paul},
TITLE = {{A Review of Semiconductor Based Ionising Radiation Sensors Used in Harsh Radiation Environments and Their Applications}},
JOURNAL = {Radiation},
VOLUME = {1},
YEAR = {2021},
NUMBER = {3},
PAGES = {194--217},

ISSN = {2673-592X},
DOI = {10.3390/radiation1030018}
}

@article{10.1063/1.4894019,
    author = {Pacella, D. and Romano, A. and Gabellieri, L. and Causa, F. and Murtas, F. and Claps, G. and Choe, W. and Lee, S. H. and Jang, S. and Jang, J. and Hong, J. and Jeon, T. and Lee, H.},
    title = {{X-ray diagnostic developments in the perspective of DEMO},
    journal = {AIP Conference Proceedings}},
    volume = {1612},
    number = {1},
    pages = {23-30},
    year = {2014},
    month = {08},
    issn = {0094-243X},
    doi = {10.1063/1.4894019},
    
}

@article{Ryan_2018,
doi = {10.1088/1748-0221/13/02/P02030},
year = {2018},
month = {feb},
publisher = {},
volume = {13},
number = {02},
pages = {P02030},
author = {Ryan, D.F. and Baumgartner, W.H. and Wilson, M. and Benmoussa, A. and Campola, M. and Christe, S.D. and Gissot, S. and Jones, L. and Newport, J. and Prydderch, M. and Richards, S. and Seller, P. and Shih, A.Y. and Thomas, S.},
title = {{Tolerance of the High Energy X-ray Imaging Technology ASIC to potentially destructive radiation processes in Earth-orbit-equivalent environments}},
journal = {Journal of Instrumentation}
}

@inproceedings{viswanathan2018quantitative,
  author    = {Viswanathan, S. and Chandrasekaran, S. and Mathiyarasu, R. and Sivasubramanian, K. and Baskaran, R. and Bhaskaran, R. and Balasubramaniam, V.},
  title     = {{Quantitative Evaluation of shielding integrity of thick structures in nuclear facilities through Radiometry}},
  booktitle = {15th Asia Pacific Conference for Non-Destructive Testing},
  year      = {2018},
  month     = nov,
  address   = {Singapore},
  note      = {Published in e-Journal of Nondestructive Testing, Vol. 23(3)},
  
}

@article{Ramli_2020,
doi = {10.1088/1757-899X/785/1/012051},
year = {2020},
month = {apr},
publisher = {IOP Publishing},
volume = {785},
number = {1},
pages = {012051},
author = {Ramli, Nurhayati and Karim, Julia Abdul and Razi, Hasniyati Md and Ariff Mustafa, Muhammad Khairul and Masenwat, Noor Azreen and Sani, Suhairy and Hussain, Mohd Huzair},
title = {{Ageing assessment of biological shielding integrity for PUSPATI TRIGA Reactor}},
journal = {IOP Conference Series: Materials Science and Engineering}
}

@Article{s23177328,
AUTHOR = {Abbene, Leonardo and Buttacavoli, Antonino and Principato, Fabio and Gerardi, Gaetano and Bettelli, Manuele and Zappettini, Andrea and Bazzi, Massimiliano and Bragadireanu, Mario and Cargnelli, Michael and Carminati, Marco and Clozza, Alberto and Deda, Griseld and Del Grande, Raffaele and De Paolis, Luca and Fabbietti, Laura and Fiorini, Carlo and Guaraldo, Carlo and Iliescu, Mihail and Iwasaki, Misahiko and Khreptak, Aleksander and Manti, Simone and Marton, Johann and Miliucci, Marco and Moskal, Pawel and Napolitano, Fabrizio and Niedźwiecki, Szymon and Ohnishi, Hiroaky and Piscicchia, Kristian and Sada, Yuta and Sgaramella, Francesco and Shi, Hexi and Silarski, Michalł and Sirghi, Diana Laura and Sirghi, Florin and Skurzok, Magdalena and Spallone, Antonio and Toho, Kairo and Tüchler, Marlene and Doce, Oton Vazquez and Yoshida, Chihiro and Zmeskal, Johannes and Scordo, Alessandro and Curceanu, Catalina},
TITLE = {Potentialities of CdZnTe Quasi-Hemispherical Detectors for Hard X-ray Spectroscopy of Kaonic Atoms at the DAΦNE Collider},
JOURNAL = {Sensors},
VOLUME = {23},
YEAR = {2023},
NUMBER = {17},
ARTICLE-NUMBER = {7328},
PubMedID = {37687783},
ISSN = {1424-8220},
DOI = {10.3390/s23177328}
}

@Article{20220205,
title = {{Estimation of the radiation backgrounds in the CEPC vertex detector}},
journal = {Radiation Detection Technology and Methods},
volume = {6},
number = {2},
pages = {170-178},
year = {2022},
issn = {2509-9930},
doi = {10.1007/s41605-022-00320-w},	
author = {Wei Xu and Haoyu Shi and Hongbo Zhu and Ke Li and Sha Bai and Xinchou Lou}
}

@article{Lee2014RadiationTF,
  title={{Radiation tests for a single-GEM-loaded gaseous detector}},
  author={Kyong Sei Lee and B. Hong and Sung Keun Park and Sang Yeol Kim},
  journal={Journal of the Korean Physical Society},
  year={2014},
  volume={65},
  pages={1367 - 1373},
}

@article{10.1063/5.0040571,
    author = {Barbui, T. and Delgado-Aparicio, L. F. and Pablant, N. and Disch, C. and Luethi, B. and Pilet, N. and Stratton, B. and VanMeter, P.},
    title = {{Multi-energy calibration of a PILATUS3 CdTe detector for hard x-ray measurements of magnetically confined fusion plasmas}},
    journal = {Review of Scientific Instruments},
    volume = {92},
    number = {2},
    pages = {023105},
    year = {2021},
    month = {02},
    issn = {0034-6748},
    doi = {10.1063/5.0040571},
}

@article{FROJDH201543,
title = {{Spectral response of the energy-binning Dosepix ASIC coupled to a 300μm silicon sensor under high fluxes of synchrotron radiation}},
journal = {Nuclear Instruments and Methods in Physics Research Section A: Accelerators, Spectrometers, Detectors and Associated Equipment},
volume = {804},
pages = {43-49},
year = {2015},
issn = {0168-9002},
doi = {https://doi.org/10.1016/j.nima.2015.09.018},
author = {E. Fröjdh and F. Bisello and M. Campbell and J. Damet and E. Hamann and T. Koenig and W.S. Wong and M. Zuber}
}

@ARTICLE{10.3389/fphy.2025.1576227,
    
AUTHOR={Tao, Luyan  and Feng, Song  and Yang, Yiwei  and Zheng, Bo },
           
TITLE={Dosimetry characteristics of ultra-high dose rate X-ray: a short review},
          
JOURNAL={Frontiers in Physics},
          
VOLUME={13},
  
YEAR={2025},
  
DOI={10.3389/fphy.2025.1576227},
  
ISSN={2296-424X}}

@article{KEALL2025787,
title = {{Real-Time Dose-Guided Radiation Therapy}},
journal = {International Journal of Radiation Oncology*Biology*Physics},
volume = {122},
number = {4},
pages = {787-801},
year = {2025},
note = {Adaptive Radiation Therapy},
issn = {0360-3016},
doi = {https://doi.org/10.1016/j.ijrobp.2025.04.019},
author = {Paul J. Keall and Issam {El Naqa} and Martin F. Fast and Emily A. Hewson and Nicholas Hindley and Per Poulsen and Chandrima Sengupta and Neelam Tyagi and David E.J. Waddington}
}

@article{FREITASNASCIMENTO2025107344,
title = {Review of real time 2D dosimetry in external radiotherapy: Advancements and techniques},
journal = {Radiation Measurements},
volume = {180},
pages = {107344},
year = {2025},
issn = {1350-4487},
doi = {https://doi.org/10.1016/j.radmeas.2024.107344},
author = {Luana de {Freitas Nascimento} and Alessia Gasparini}
}

@article{https://doi.org/10.1118/1.4926282,
author = {Keall, P.},
title = {{TH-D-BRD-04: A Review of Present and Near-Future Methods for Real-Time Target Tracking and Adaptation}},
journal = {Medical Physics},
volume = {42},
number = {6Part44},
pages = {3738-3739},
doi = {https://doi.org/10.1118/1.4926282},
year = {2015}
}

@article{Jegal2024MotionMA,
  title={{Motion Management and Image-Guided Technique in Photon Radiation Therapy: A Review of an Advanced Technology}},
  author={Jin Jegal and Hyojun Park and Seonghee Kang and Chang Heon Choi and Jung-in Kim},
  journal={Progress in Medical Physics},
  year={2024},
}

@article{PALMERINI2002159,
title = {{Design of the radiation shielding for a microsatellite}},
journal = {Acta Astronautica},
volume = {50},
number = {3},
pages = {159-166},
year = {2002},
issn = {0094-5765},
doi = {https://doi.org/10.1016/S0094-5765(01)00151-5},
author = {Giovanni B Palmerini and Francesco Pizzirani}
}

@article{Zhang_2011,
doi = {10.1088/1748-0221/6/11/C11013},
year = {2011},
month = {nov},
publisher = {},
volume = {6},
number = {11},
pages = {C11013},
author = {J Zhang and E Fretwurst and R Klanner and H Perrey and I Pintilie and T Poehlsen and J Schwandt},
title = {{Study of X-ray radiation damage in silicon sensors}},
journal = {Journal of Instrumentation}
}

@article{SINGH2006713,
title = {{Reliability and performance limitations in SiC power devices},
journal = {Microelectronics Reliability}},
volume = {46},
number = {5},
pages = {713-730},
year = {2006},
issn = {0026-2714},
doi = {https://doi.org/10.1016/j.microrel.2005.10.013},
author = {Ranbir Singh}
}

@article{COWEN201873,
title = {{Point defects production and energy thresholds for displacements in crystalline and amorphous SiC}},
journal = {Computational Materials Science},
volume = {151},
pages = {73-83},
year = {2018},
issn = {0927-0256},
doi = {https://doi.org/10.1016/j.commatsci.2018.04.063},

author = {Benjamin J. Cowen and Mohamed S. El-Genk},

}

@ARTICLE{4033220,
  author={Bellini, Marco and Jun, Bongim and Chen, Tianbing and Cressler, John D. and Marshall, Paul W. and Chen, Dakai and Schrimpf, Ronald D. and Fleetwood, Daniel M. and Cai, Jin},
  journal={IEEE Transactions on Nuclear Science}, 
  title={{X-Ray Irradiation and Bias Effects in Fully-Depleted and Partially-Depleted SiGe HBTs Fabricated on CMOS-Compatible SOI}}, 
  year={2006},
  volume={53},
  number={6},
  pages={3182-3186},
  doi={10.1109/TNS.2006.885795}}

@article{10.1063/5.0179556,
    author = {Migliore, F. and Alessi, A. and Principato, F. and Girard, S. and Cannas, M. and Gelardi, F. M. and Lombardo, A. and Vecchio, D. and Brischetto, A. and Agnello, S.},
    title = {{$\beta$-rays induced displacement damage on epitaxial 4H-SiC revealed by exciton recombination}},
    journal = {Applied Physics Letters},
    volume = {124},
    number = {4},
    pages = {042101},
    year = {2024},
    month = {01},
}

@ARTICLE{11303914,
  author={Yang, Tao and Satapathy, Yashas and Sekely, Ben J. and Tishelman-Charny, Abraham and Allion, Greg and Atar, Gil and Barletta, Philip and Haber, Carl and Holland, Steve and Muth, John F. and Pavlidis, Spyridon and Stucci, Stefania},
  journal={IEEE Transactions on Nuclear Science}, 
  title={{Time Resolution Characterization of 4H-SiC LGADs With a 90Sr Source}}, 
  year={2026},
  volume={73},
  number={2},
  pages={407-411},
  doi={10.1109/TNS.2025.3646114}}

@article{10.1063/1.364397,
    author = {Hemmingsson, C. and Son, N. T. and Kordina, O. and Bergman, J. P. and Janzén, E. and Lindström, J. L. and Savage, S. and Nordell, N.},
    title = {{Deep level defects in electron-irradiated 4H SiC epitaxial layers}},
    journal = {Journal of Applied Physics},
    volume = {81},
    number = {9},
    pages = {6155-6159},
    year = {1997},
    month = {05},
    issn = {0021-8979},
    doi = {10.1063/1.364397},
    

}

\end{document}